\documentclass[%
 aip,
 jmp,%
 amsmath,amssymb,
 reprint,%
]{revtex4-1}

\usepackage{graphicx}
\usepackage{dcolumn}
\usepackage{bm}
\usepackage{mathtools}
\usepackage{hyperref}
\usepackage{mathrsfs}

\hypersetup{
    colorlinks,
    citecolor=black,
    filecolor=black,
    linkcolor=black,
    urlcolor=black
}

\begin{document}


\title[The Colored Noise of Magnetic Monopoles: Subdiffusion  in a Coevolving Vacuum  and Spin Ice Exponents]{The Colored Noise of Magnetic Monopoles: Subdiffusion  in a Coevolving Vacuum  and Spin Ice Exponents}

\author{Cristiano Nisoli}

\affiliation{ 
Theoretical Division, Los Alamos National Laboratory, Los Alamos, NM, 87545, USA
}%


\date{\today}

\begin{abstract}
We relate the anomalous noise found experimentally in spin ice  to the subdiffusion  of magnetic monopoles. Because monopoles are emergent particles, they do not move in a structureless vacuum. Rather, the underlying spin ensemble filters the thermal white noise, leading to non-trivial coevolution. Thus, monopoles can be considered as  random walkers under the effect of stochastic forces only as long as those are not trivially white, but instead subsume the evolution of the spin vacuum.  Via this conceptualization, we conjecture relations between the color of the noise and other observables, such as relaxation time, monopole density, the dynamic exponent, and the order of the annihilation reaction, which lead us to introduce spin ice specific critical exponents in a neighborhood of the ice manifold.
\end{abstract}

\maketitle

Spin ice materials~\cite{Ramirez1999,harris1997geometrical,bramwell2020history,Nisoli2013colloquium,skjaervo2019advances,ortiz2019colloquium} 
are exotic magnets that interrogate notions of order, constrained disorder~\cite{henley2010coulomb}, classical topological order~\cite{macdonald2011classical,henley2011classical,lao2018classical}, Pauling entropy~\cite{Pauling1935}, frustration, and emergence. 
In their crystalline forms, they reveal  magnetic monopoles~\cite{ryzhkin2005magnetic2,Castelnovo2008} as emergent particles describing strongly correlated local violations of their topological structure~\cite{castelnovo2012spin}. 

Recently, a  series of carefully performed experiments measured magnetic noise in spin ice~\cite{dusad2019magnetic,goryca2021field} (see also previous works on dynamical susceptibilities~\cite{bovo2013brownian,ehlers2002dynamical}), and related it to monopole dynamics. 
The experimental  spectral power is generally well fitted~\footnote{Or also
 $P(\omega)\propto {1}/({\omega_0^{\beta}+\omega^{\beta}})$  used by ref~\cite{dusad2019magnetic}. The two expressions are asymptotic at low and high frequencies, the regimes that interest us.} by
\begin{equation}
P(\omega)\propto(\omega_0^2+\omega^2)^{-\beta/2}.
\label{PA}
\end{equation}
Clearly,  $\tau_0=2\pi/\omega_0$ is the relaxation time, whereas $\beta$ is the ``color'' of the noise, found experimentally to be between pink and brown ($1<\beta<2$). 

The introduction of spin noise spectroscopy to the study of spin ice materials, natural or artificial, is a most welcomed development. Indeed,  as we will show in a future, broader work on more general~\cite{Morrison2013,nisoli2017deliberate} geometries, a number of relevant observables can be extracted from the power spectral density, such as the total rate of spin flips, as
\begin{equation} 
\nu \simeq \frac{\tau}{8\pi}\int_0^{\frac{2\pi}{\tau}} \!\! P(\omega)\omega^2 d\omega,
\label{nu}
\end{equation}
where the ultraviolet cutoff $\tau$ 
 is the time discretization of the measure~\cite{0034-4885-79-10-106501}. It is chosen to be small enough that spin flips in $\tau$ are negligible, but not as small as to record the internal processes of magnetic moment inversion.  
 Remarkably, it might be possible to extract the spin ice part of the specific heat and thus the magnetic part of entropy from the magnetic fluctuations. This would be most important in the case of nano-magnetic realizations, where the magnetic part of the specific heat is many orders of magnitude smaller than the phonon term, and thus impossible to detect via calorimetric measures. 
 
 Here, however, we concentrate specifically on the noise of monopoles. 
In an initially unnoticed calculation, Ryzhkin~\cite{ryzhkin2005magnetic2} introduced  spin ice monopoles precisely to compute relaxations and susceptibilities.  A subsequent work related $\tau_0$ to the creation-annihilation rates of monopole pairs, linearized around equilibrium~\cite{klyuev2017statistics}, predicting a Lorentzian power spectrum, or $\beta=2$ in (\ref{PA}), i.e.\ brown (or Brownian) noise.  

In general, recent studies have concentrated on relaxation times, while  the anomalous color of the noise ($\beta\ne2$) has perhaps not been properly appreciated. To color the noise, Klyuev {\em et al.}~\cite{klyuev2019memory}  had postulated a continuous superposition of different Brownian spectra. 
The idea of obtaining colored noise  from integrating Lorentzians over a continuum kernel  dates to 1937, to explain pink noise in triodes~\cite{bernamont1937fluctuations}. It can be  useful when such kernel is associated to some significant density of states of corresponding and understood real processes. 
Other approaches have appeared since~\cite{saletti1986comparison,kaulakys2005point}, tying colored noise to memory, incremental (anti)correlations, {\it fractional} brownian motions~\cite{mandelbrot1968fractional,kou2004generalized}, and anomalous diffusion 
for a broad phenomenology of random walks on disordered substrates, percolation clusters, fractals, et cetera ~\cite{klafter1987stochastic,metzler2000random,bouchaud1990anomalous,havlin1987diffusion}, often  
 described via the fractional Langevin equation~\cite{kobelev2000fractional,lutz2012fractional,sokolov2002fractional}, and the corresponding fractional Fokker-Planck equation~\cite{kolwankar1998local}. 

Crucially, monopoles in  pyrochlore spin ice or  square artificial spin ice~\cite{perrin2016extensive,farhan2019emergent,ostman2018interaction,Wang2006,Porro2013,Zhang2013} are emergent particles in a vacuum of constrained disorder. It is known that at equilibrium their ice rule obeying spin-vacuum induces  entropic  interactions~\cite{castelnovo2011debye,nisoli2020equilibrium,nisoli2020concept}.  It is thus hardly surprising that  their random walk is also non-trivial: their passage modifies the spin ensemble, which in turn affects their walk, leading to short time memory effects  that are lost at larger times, as equilibration ensues. 

The microscopic mechanisms can be complicated and might involve system-specific hidden variables~\cite{tomasello2019correlated,stern2019quantum,gliga2015broken}. Indeed, Monte Carlo simulations---otherwise most useful in describing these systems at equilibrium---have generally failed to reproduce the proper value of $\beta$~\cite{dusad2019magnetic,goryca2021field}. 

Because of these specificities, in the following we will attempt a phenomenological conceptualization to squeeze heuristic informations from the color of the noise. Some considerable amount of hand-waving will necessarily be involved, for which we apologize in advance. 

Consider the  magnetization signal  as $M(t)= \langle M \rangle+ 2m(t)$ recorded along a direction. The fluctuation $m$ is a stochastic process and can always be written as
\begin{equation}
m(t) =  \int^t_{-\infty} \!\!\!\!\! \chi(t-s)\eta(s)ds.
\label{m}
\end{equation}
where $\eta$ is a white noise which describes uncorrelated thermal fluctuations: $\langle \eta(t) \rangle =0$,  $\langle \eta(t) \eta(t')\rangle =\overline{\eta^2} \delta(t-t')$, and $\overline{\eta^2} \propto T$ where $T$ is the temperature. $\chi$ is the susceptibility kernel that ``filters'' the noise.
From (\ref{m}), the mean squared displacement of the signal is 
\begin{equation}
\sigma_m^2(t)=\langle [m(t)-m(0)]^2\rangle= \overline {\eta^2} \int_0^t \!\! \chi(s)^2 ds.
\label{sigmam}
\end{equation}
In frequency domain, from (\ref{m}) we have $\tilde m(\omega) =\tilde \chi_{_{\theta}} (\omega) \tilde \eta(\omega)$  where $\chi_{_{\theta}}(t)=\theta(t)\chi(t)$ and $\theta(t)$ is the Heaviside step. Thus, the power spectral density is given by
\begin{equation}
P(\omega) =\langle |\tilde m(\omega)|^2 \rangle = \overline{\eta^2} |\tilde \chi_{_{\theta}}(\omega)|^2.
\label{P2}
\end{equation}

 In pyrochlore and square spin ice at low $T$, the magnetic noise at high frequency is proportional to the walk of monopoles. A spin flip either: creates/annihilates a monopole pair, an inelastic process associated to relaxation and the lower end of the frequency spectrum; or changes a monopole position, an elastic process pertaining to high frequencies. 
 
 Calling $\vec x(t)$ the position of the monopole, a hop $\Delta \vec x$ causes a  change in magnetization $\delta M_i= \pm 2\Delta x_i$. 
Thus,  the monopole's walk  obeys 
\begin{equation}
\frac{d \vec x }{dt} = \vec f (t)
\label{L}
\end{equation}
where  $\vec f$ is the stochastic force   given, from (\ref{m}), by
\begin{equation}
\vec f(t)=\int^t_{-\infty} \!\!\!\!\! \Pi(t-s) \vec\eta(s)ds.
\label{f}
\end{equation}
Which can be interpreted as such: the thermal white noise  {$\vec \eta$ [$\langle \eta_i(t) \eta_j(t')\rangle =\overline{\eta^2} \delta_{ij} \delta(t-t')$]}
is in general filtered by the kernel
\hbox{$\Pi(t)=\delta(t)+{d \chi}/{dt}$}, an effect that must come from the spin vacuum. If $\chi$ is constant, monopoles perform a trivial Brownian walk, associated to diffusion and brown noise.  

For definiteness,  we make an ansatz on $\chi$. The choice  
\begin{equation}
\chi (t) \propto \left( \frac{\tau}{t} \right)^{1-\beta/2} \!\!\! e^{-\omega_0 t}
\label{chi}
\end{equation}
returns~\footnote{from (\ref{P2}) and $\int_0^{\infty} \chi(t) \exp(i\omega t)dt=\bar{t}^{1-\beta/2}\Gamma (\beta/2)/(\omega_0-i \omega)^{\beta/2}$ for $\beta>0$.} the functional form  of (\ref{PA}).
Because the relaxation time diverges at small $T$, and because we are interested  in  the color of the noise, we  take from now on the approximation $\omega_0\simeq0$. Our claims  will pertain to timescales much shorter than relaxation, which therefore do not correspond to ergodicity. In the following equations, ultraviolet divergences are cured by the cutoff of the Nyquist frequency $2\pi/\tau$ (or at small times by $\tau$). 

Equation~(\ref{chi}) tells us that when $\beta\ne2$ there is an algebraically decaying memory for fast processes, so that (\ref{f}) becomes
\begin{equation}
\vec f = \vec \eta (t) - (1-\beta/2)\int_{-\infty}^{t} \!\!\!\!\! \vec \eta(s) (t-s)^{\beta/2-2}  ds:
\label{L2}
\end{equation}
along with the thermal fluctuations at equal time [first term in (\ref{L2})], the monopole experiences an extra force reminiscent of past fluctuations, opposite in direction to previous fluctuations  [second term in (\ref{L2}), which disappears when $\beta=2$.].  Similarly,
\begin{equation}
\vec x(t) =  \int^t_{-\infty} \!\!\!\!\! \vec \eta(s) (t-s)^{\beta/2-1} ds,
\label{x}
\end{equation}
which returns a Brownian motion when $\beta=2$.

From (\ref{sigmam}), (\ref{x})  the mean square displacements of the monopoles are 
\begin{align}
\sigma_{x_i}^2&= \frac {\overline{\eta ^2} \tau}{\beta-1} \left[(t/\tau)^{\beta-1}-1\right]~ &\text{for}~\beta\ne 1 \nonumber \\
\sigma_{x_i}^2&= \overline{\eta ^2}{\tau} \ln(t/\tau)~
&\text{for}~\beta =1
\label{sigmam2}
\end{align}
%
With similar gymnastics one obtains  the correlation among successive increments $\delta x_i\simeq \tau \frac{dx_i}{dt}$  as
\begin{align}
\langle \delta x_i(t) \delta x_j(0) \rangle 
= \delta_{ij}  \overline{\eta ^2} \tau (\beta/2-1) (t/\tau)^{\beta/2-2}.
\label{incr}
\end{align}
Equations (\ref{sigmam2}) follow straightly from the fact that $\chi$ is algebraic when $\omega_0\simeq 0$. Indeed, from (\ref{x})  
one finds the scaling
\begin{equation}
\langle x_{i}(\lambda t)   x_{j}(\lambda t') \rangle = \delta_{ij} \lambda^{\beta-1} \langle  x_{i}(t)  x_{j}( t') \rangle
\label{affine}
\end{equation}
showing that the process is self-affine. Equations (\ref{sigmam2}) can be also obtained from (\ref{affine}).

From (\ref{sigmam2}, \ref{incr}), $\beta$ controls different regimes. 

Because of  { \em positive} correlations between successive increments, for $\beta>3$ the process is superdiffusive ($\sigma^2/t \to \infty$), ballistically so ($\sigma \propto t$) for $\beta=3$ and hyper-ballistic ($\sigma/t \uparrow +\infty$) for \hbox{$\beta>3$}. 

For $\beta=2$ (brown noise) the process is {\em diffusive} (\hbox{$\sigma^2/t \sim 1$}) because increments are {\em uncorrelated} (Brownian  walk). 

For $1<\beta<2$ (pink-brown noise interval) the process is {\it subdiffusive} ($\sigma$ diverges but $\sigma^2/t \to 0$) because increments are {\em anti-correlated}. 

For $\beta=1$ (pink noise) increments {\em anti-correlate} strongly enough to turn the diffusion from algebraic to logarithmic ($\sigma^2 \! \sim \ln t$). We might call this state {\em critically subdiffusive} as it separates subdiffusion from non-diffusion (and might explain the ubiquity of pink noise in complex systems, especially biological, believed by some to be poised in the neighborhood of criticality~\cite{mora2011biological}).

For $\beta<1$ there is no diffusion and $\sigma^2$ converges to $\overline{\eta ^2}/(1- \beta)$. 
 
Repetita iuvant: we have taken $\tau \omega_0 \simeq 0$, which means that the classification above is only true for timescales small with respect of relaxation and equilibration. Note also that $\beta$ is related to the  Hurst exponent~\cite{hurst1956methods,carbone2004time} $H$,  a measure of long-term memory that controls the autocorrelation of increments in time series via $H=(\beta-1)/2$.


In  pyrochlore ice, Dusad {\it et al.} found~\cite{dusad2019magnetic} values  $\beta\simeq 1.2-1.5$, suggesting that monopoles subdiffuse. Preliminary unpublished data suggest subdiffusion in degenerate square ice also~\footnote{Personal communications with Michael Saccone on data from A. Farhan's group.}.  The case of non-degenerate square ice is tricky because the degeneracy is lifted, and we shall consider it elsewhere. 

As already discussed, the origin of the anticorrelation of increments can be system-specific and involves short time processes not easily captured by a Monte Carlo simulation. A purely geometrical anticorrelation exists, but accounts for a very small deviation from brown noise. 
At low $T$,  at each hop only $3$ out of $4$ spins surrounding the monopole can flip without creating a $\pm 4$ charge; so that, when a monopole enters a vertex, its next hop has memory of its previous one.

One can estimate the effect of the purely geometric anticorrelation. If $p$ is the probability that the monopoles hops, 
the anti-correlation of consecutive hops among tetrahedra  in pyrochlores is
%
 $\langle \delta {\bf x}' \cdot \delta {\bf x} \rangle=-p/9$. 
%
For degenerate square ice, the anticorrelation is zero if the vertex was of the Type-II~\cite{Wang2006} kind before the monopole entered it. If it was  of the Type-I kind we have
%
 $\langle \delta {\bf x}' \cdot \delta {\bf x} \rangle=-p/3$. 
Because at low temperature Type-I vertices appear with a frequency $\simeq 0.38$ in a pure six-vertex model, we find in two dimensions an {\em average} anticorrelation strength $\simeq  0.127 p$ not far from the value $0.\bar1 p$ of the 3D case. 
If from (\ref{affine}) we consider the monopole walk as a self-affine process, we can venture to guess $\beta_g$ resulting from purely geometric  constraints. If we assume  $\tau \overline{\eta^2}=p$ in (\ref{incr}) we have $\langle \vec \delta x(\tau) \cdot \vec \delta x(0) \rangle =  d (\beta/2-1)p$, where $d$ is the dimension ($d=2,3$). Equating it with the geometric anticorrelations of increments computed above we obtain $\beta_{g,2D}\simeq 1.87$,  $\beta_{g,3D}\simeq 1.93$. 

Experimental results for the noise of degenerate square ice have not been reported. A Glauber dynamics on the pure vertex-model (to be reported elsewhere) returns $\beta \simeq 1.9$ close to our estimate. (We expect values for real systems to be considerably lower.)

For pyrochlore spin ice, Fig.~4e in Ref.~\cite{dusad2019magnetic} shows $\beta$ getting close to 2 at low $T$ for simulations of a pure vertex model (there called NNSI), in agreement with our estimate for a purely geometrical anticorrelation. By adding dipolar interactions they find $\beta~\simeq 1.8$, suggesting a strengthening of anticorrelation of increments via long-range interaction. An even lower $\beta\simeq1.5$ is found in experiments, suggesting the presence in the real material of other sources of anticorrelation, that a Monte Carlo simulation cannot capture. 
Clearly, both long range interactions and local effects related to moment inversion can play a role in $\beta$.  

We conclude this section by mentioning that, as the reader familiar with fractional calculus~\cite{baleanu2011fractional} has surely noticed, the integrals in  (\ref{L2}, \ref{x}) are fractional. In fact, (\ref{x}) leads to the same noise spectral power of the  fractional Langevin equation
\begin{equation}
\frac{d^{\beta/2}}{dt^{\beta/2}}\vec x =  \vec{\eta}
\label{fractional}
\end{equation}
where ${d^{\beta/2}}/{dt^{\beta/2}}$ is a fractional derivative  of order $\beta/2$. Indeed, more generally, 
 our treatment is equivalent to the generalized Langevin equation
\begin{align}
&\int_{-\infty}^t \!\!\!\!\!\! \psi(t-s) \frac{d\vec x (s)}{ds}ds=
-\!\! \int_{-\infty}^t \!\!\!\!\!\! \gamma(t-s) \vec x(s)+\!\! \int^t_{-\infty} \!\!\!\!\!\! \zeta(t-s) \vec\eta(s)ds
\label{full}
\end{align}
as long as $\psi, \gamma, \zeta$ obey the constraint
\begin{equation}
{\tilde \zeta_{\theta}(\omega)}=\tilde \chi_{\theta}(\omega) \left[i\omega \tilde \psi_{\theta}(\omega)+\tilde \gamma_{\theta}(\omega)\right].
\label{constraint}
\end{equation}
In (\ref{full}), $\psi$ can describe trapping~\cite{metzler2004restaurant}, i.e.\ a non-uniformity in hopping time, $\gamma$ is a memory kernel remembering previous positions, while $\zeta$ remembers previous thermal fluctuations. Thus, the problem has two gauges, and our treatment above corresponds to $\tilde \gamma =0, \tilde \psi =1$, whereas (\ref{fractional}) corresponds to $\tilde \gamma =0, \tilde \zeta =1$, and thus from (\ref{constraint}), $\psi=(-i\omega)^{\beta/2-1}$, the kernel of a fractional derivative. 

From the conceptualization above, we submit  to future numerical and experimental analysis heuristic conjectures on the relations between the color of the noise and the exponents for the relaxation time $\tau_0$, the correlation length $\xi_0$ and rate of monopole annihilation $\nu_a$. 

 In the theory of phase transitions, critical exponents relate the algebraic divergence of various quantities to $|T-T_c|$ where $T_c$ is a critical temperature. In pyrochlore or degenerate square spin ice the ice-rule manifold at $T=0$ is a critical phase, with algebraic correlations~\cite{henley2010coulomb}. Because of that, the divergence of the correlation length $\xi_0$ and relaxation time $1/\omega_0$ are not algebraic when $T\downarrow 0$. The ice manifold is characterized by the absence of monopoles, and thus $n_m$, the density of monopole per unit vertex, is a good parameter to characterize divergences. Our arguments (below) lead us to introduce the  exponents $\theta$, $\phi$, specific to spin ice:
\begin{align}
\tau_0 & \sim n_m^{-\phi} \nonumber \\
\xi_0 & \sim n_m^{-\theta} \nonumber \\
\nu_a& \sim  n_m^x, 
\label{exponents}
\end{align}
for $n_m\downarrow 0$, in reference to steady states, not necessarily at equilibrium. If $z$ is the dynamic exponent, defined by $\tau_0 \sim \xi_0^z$, then
\begin{equation}
z=\phi/\theta.
\label{z}
\end{equation}

We now attempt to relate these exponents to the color of the noise. 
Because the relaxation time is connected to inelastic processes of monopole creation-annihilation, we begin by treating the latter as a chemical reaction
\begin{equation}
\{q=+2\} + \{q=-2\} \leftrightharpoons 2 \{q=0\}
\label{chemistry}
\end{equation}
where $\{q=\pm 2\}$ denotes a positive/negative monopole vertex and $\{q= 0\}$ an ice rule obeying vertex.  From elementary chemistry, the total rate of the reaction $\nu$ is  
\begin{equation}
\nu = \nu_a -\nu_c
\end{equation}
where $\nu_a, \nu_c$ are the annihilation and creation rates for a monopole pair, such that $\dot n_m=-2 \nu$. If $n_+, n_-$ are the densities of vertices of positive and negative charge, with $n_+=n_-=n_m/2$,  then we have from elementary chemical kinetics
\begin{align}
\nu_a&=k_a n_m^x \nonumber \\
\nu_c&=k_c (1-n_m)^2
\end{align}
where $k_a, k_c$ are the temperature dependent rate constants, and $x$ is called the order of the (annihilation) reaction. 

In diffusive regime, the elementary annihilation reaction has an order $x=2$. 
Our case, instead, is one of subdiffusion~\footnote{Furthermore the reaction is not necessarily elementary, but might involve intermediate steps from so-called incontractible pairs~\cite{castelnovo2010thermal}, though those are infrequent in absence of quenches.}, suggesting \hbox{$x\ne2$}.

Because an activation energy is needed to create monopoles, whereas $\nu_a$ is related to the hoppng frequency, it must be that $k_c/k_a \downarrow 0$ at low $T$. The steady state (which is {\it not necessarily} an equilibrium state) is found for $\nu=0$ which returns $n_m\simeq (k_c/k_a)^{1/x}$ at lowest order in $k_c$. Then, from the  ubiquitously employed linearization $\omega_0=d\nu/dn_m$ (computed at the steady state), simple algebra shows
\begin{equation}
\omega_0= x k_a n_m^{x-1} +O(n_m^x),
\label{ciccia}
\end{equation}
which implies 
\begin{equation}
\phi=x-1.
\label{phi}
\end{equation}

To relate $x$ and thus $\phi$ to $\beta$ further congectures are needed. The study of reactions in subdiffusive regimes is recent, generous of sophisticated mathematical methods, yet stil orphan of a widely accepted paradigm~\cite{yuste2002subdiffusion,sokolov2002fractional,sokolov2006reaction}. As those  issues are beyond the scope of the present work,  we try instead to get away with a heuristic  trick.

The rate of annihilation for a single positive monopoles is $\nu_a/n_+$. In diffusive regime \hbox{$x=2$} and thus \hbox{$\nu_a/n_+ \propto k_a n_-$}. 
Introducing the time $\tau_-$ to diffuse along the average distance among monopoles, $l_-=n_-^{-d}$ ($d$ is the dimension) we have then $\nu_a/n_+ \propto k_a \tau_-^{-d/2}$. This is intuitive: the likelihood that a positive monopole finds a negative one decreases as the diffusion time to cover their average distance increases, and it also decreases exponentially with increased dimensionality, since at higher dimensions it is less likely for random walkers to find each others. We now declare our sleight of hand: we account for subdiffusion by computing $\tau_-$ from (\ref{sigmam2}) as $\tau_- \sim l_-^{2/(\beta-1)}$. We obtain therefore $\nu_a/n_+ \propto k_a n_-^{1/(\beta-1)}$, and remembering that $n_+=n_-=n_m/2$ we find
\begin{equation}
x={\frac{\beta}{(\beta-1)}}.
\label{xchem}
\end{equation}
Thus
from (\ref{xchem}) we have
\begin{equation}
\phi={1/{(\beta-1)}},
\label{phi2}
\end{equation}
which relates the exponent for the divergence of relaxation time to the color of the noise. Note that both $\phi$ and $x$ diverge at $\beta=1$, corresponding to the already mentioned subdiffusive criticality. From (\ref{phi2}) and (\ref{z}) we have then for the dynamic exponent
%
\begin{equation}
z= {\theta^{-1}}{(\beta-1)^{-1}}
\end{equation}
Via a mean field approximation~\cite{nisoli2020equilibrium}  {\it at equilibrium} we had found $\xi^2\sim 1/\overline{q^2}\sim 1/n_m$  ($\overline{q^2} =4 n_m$ is the average charge per vertex) which would imply $\theta=1/2$, $z=2/(\beta-1)$. 
Interestingly,  Chen and Yu~\cite{chen2007measurement} found for the 2D Ising model an analogous result via completely different scaling arguments. They obtained $z = 1.75/(\beta-1)$.

Finally, in quenches to very low temperature, where we can assume $n_m\sim 0, k_c\sim 0, \omega_0\sim 0$ in the steady state, the rate equation becomes $\dot n_m=-2\nu \simeq -2k_a n^x$ and thus $n_m$ decays algebraically as 
\begin{equation}
n_m\sim t^{-1/\phi}=t^{1-\beta}
\end{equation}
which reduces to the decay $n_m\sim 1/t$ in the Brownian case, a well known result~\cite{}.

We have attempted to contextualize the anomalous exponent in the noise of spin ice materials within the framework of anomalous diffusion, in the context of a random walk of monopoles as emergent particles in a non-trivial vacuum. The microscopic details of the mechanism is certainly system-specific.  In Ho$_2$Ti$_2$O$_7$ and Dy$_2$Ti$_2$O$_7$, it can involve, for instance, correlated quantum tunneling~\cite{tomasello2019correlated,stern2019quantum} among other effects. In nano realizations, where magnetic moments are not truly binary variables but rather composite mesoscopic objects, the microscopic mechanism of moment inversion in nanoislands  can be specific and its activation energy can depend on the relaxation of magnetization at the tip, itself resulting from collective interactions~\cite{gliga2015broken}. Similarly, in nanowires network it can depend on domain walls in the vertices~\cite{daunheimer2011reducing,burn2017dynamic}. We have proposed spin ice specific critical exponents and related them to the color of the noise, which interrogates future experimental and theoretical work.

 This work was carried out under the auspices of the U.S.
DoE through the Los Alamos National
Laboratory, operated by
Triad National Security, LLC
(Contract No. 892333218NCA000001) and founded by a grant from the DOE-LDRD office at Los Alamos National
Laboratory.

\bibliography{library2.bib}{}

\end{document}